\documentclass{PoS}
\usepackage{lineno}

\title{Chiral phase transition in (2 + 1)-flavor QCD}

\ShortTitle{Chiral phase transition in (2 + 1)-flavor QCD}
\author{ 
	Heng-Tong Ding$^1$, Prasad Hegde$^2$, Olaf Kaczmarek$^{1,3}$, Frithjof Karsch$^{3,4}$, Anirban Lahiri$^3$, \speaker{Sheng-Tai Li}$^1$\thanks{This work was supported in part by
		the Deutsche Forschungsgemeinschaft (DFG) through the grant 315477589-TRR 211,
		the grants 05P15PBCAA and 05P18PBCA1 of the German Bundesministerium f\"ur Bildung und
		Forschung, the National Natural
		Science Foundation of China under grant numbers 11775096 and 11535012,
		Furthermore, this work was supported through Contract No.~DE-SC0012704 with the
		U.S. Department of Energy, through the Scientific Discovery through Advanced
		Computing (SciDAC) program funded by the U.S. Department of Energy, Office of
		Science, Advanced Scientific Computing Research and Nuclear Physics and
		the DOE Office of Nuclear Physics funded BEST topical collaboration, 
		and a Early Career Research Award of the Science and Engineering
		Research Board of the Government of India.
		Numerical calculations have been made possible through PRACE grants
		at CSCS, Switzerland, and at CINECA, Italy as well as grants at the
		Gauss Centre for Supercomputing and 
		NIC-J\"ulich, Germany. These grants provided access to resources on
		Piz Daint at CSCS, Marconi at CINECA as well as on 
		JUQUEEN and JUWELS at NIC.
		Additional calculations have been performed on 
		GPU clusters of USQCD, 
		at Bielefeld University, the PC$^2$ Paderborn 
		University and the Nuclear Science Computing Center at Central China
		Normal University, Wuhan, China. Some data sets have also partly been
		produced at the TianHe II Supercomputing Center in Guangzhou.}, Swagato Mukherjee$^{4}$, Peter Petreczky$^{4}$ (for the HotQCD collaboration) \\
	\llap{$^1$}	Key Laboratory of Quark \& Lepton Physics (MOE) and Institute of Particle Physics,  \\
	Central China Normal University, Wuhan 430079, China. \\
	\llap{$^2$}	Center for High Energy Physics, Indian Institute of Science, Bangalore 560012, India\\	
	\llap{$^3$}	Fakult\"at f\"ur Physik, Universit\"at Bielefeld, D-33615 Bielefeld, Germany\\
	\llap{$^4$} Physics Department, Brookhaven National Laboratory, Upton, NY 11973, USA\\
E-mail: 
\email{hengtong.ding@mail.ccnu.edu.cn},
\email{prasadhegde@iisc.ac.in},
\email{okacz@physik.uni-bielefeld.de},
\email{karsch@physik.uni-bielefeld.de},
\email{alahiri@physik.uni-bielefeld.de},
\email{lishengtai@mails.ccnu.edu.cn},
\email{swagato@bnl.gov},
\email{petreczk@quark.phy.bnl.gov}

	}

\abstract{The chiral phase transition temperature $T_{c}^{0}$ is a fundamental quantity of QCD. To determine this quantity we have performed simulations of (2 + 1)-flavor QCD using the Highly Improved Staggered Quarks (HISQ/tree) action on $N_{\tau}=6, 8$ and 12 lattices with aspect ratios $N_{\sigma}/N_{\tau}$ ranging from 4 to 8. 
	In our simulations the strange quark mass is fixed to its physical value $m_{s}^{\rm{phy}}$, and the values of two degenerate light quark masses $m_{l}$ are varied from $m_{s}^{\rm{phy}}/20$ to $m_{s}^{\rm{phy}}/160$ which correspond to a Goldstone pion mass $m_{\pi}$ ranging from 160 MeV to 55 MeV in the continuum limit.
	By investigating the light quark mass dependence and the volume dependence of various chiral observables, e.g. chiral susceptibilities and Binder cumulants, no evidence for a first order phase transition in our current quark mass window is found.
	Two estimators $T_{60}$ and $T_{\delta}$ are proposed to extract the chiral phase transition temperature $T_{c}^{0}$ in the chiral and continuum limit and our current estimate for $T_{c}^{0}$ is $132_{-6}^{+3}$ MeV.

	}

\FullConference{The 36th Annual International Symposium on Lattice Field Theory - LATTICE2018\\
		22-28 July, 2018\\
		Michigan State University, East Lansing, Michigan, USA.}

\begin{document}
\section{Introduction}
\begin{figure}[htp]
	\begin{center}
		\includegraphics[width=0.35\textwidth]{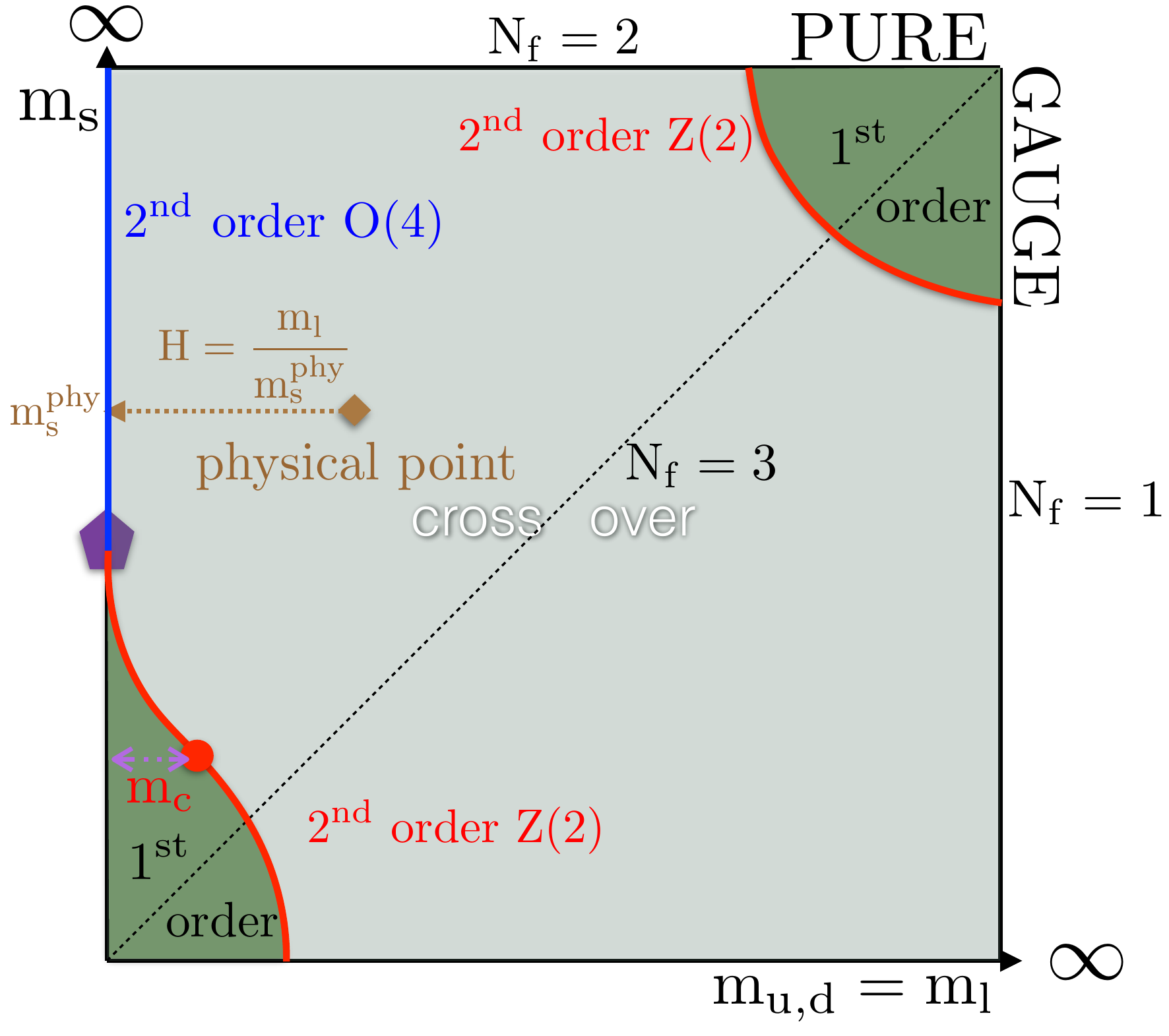}
		\includegraphics[width=0.35\textwidth]{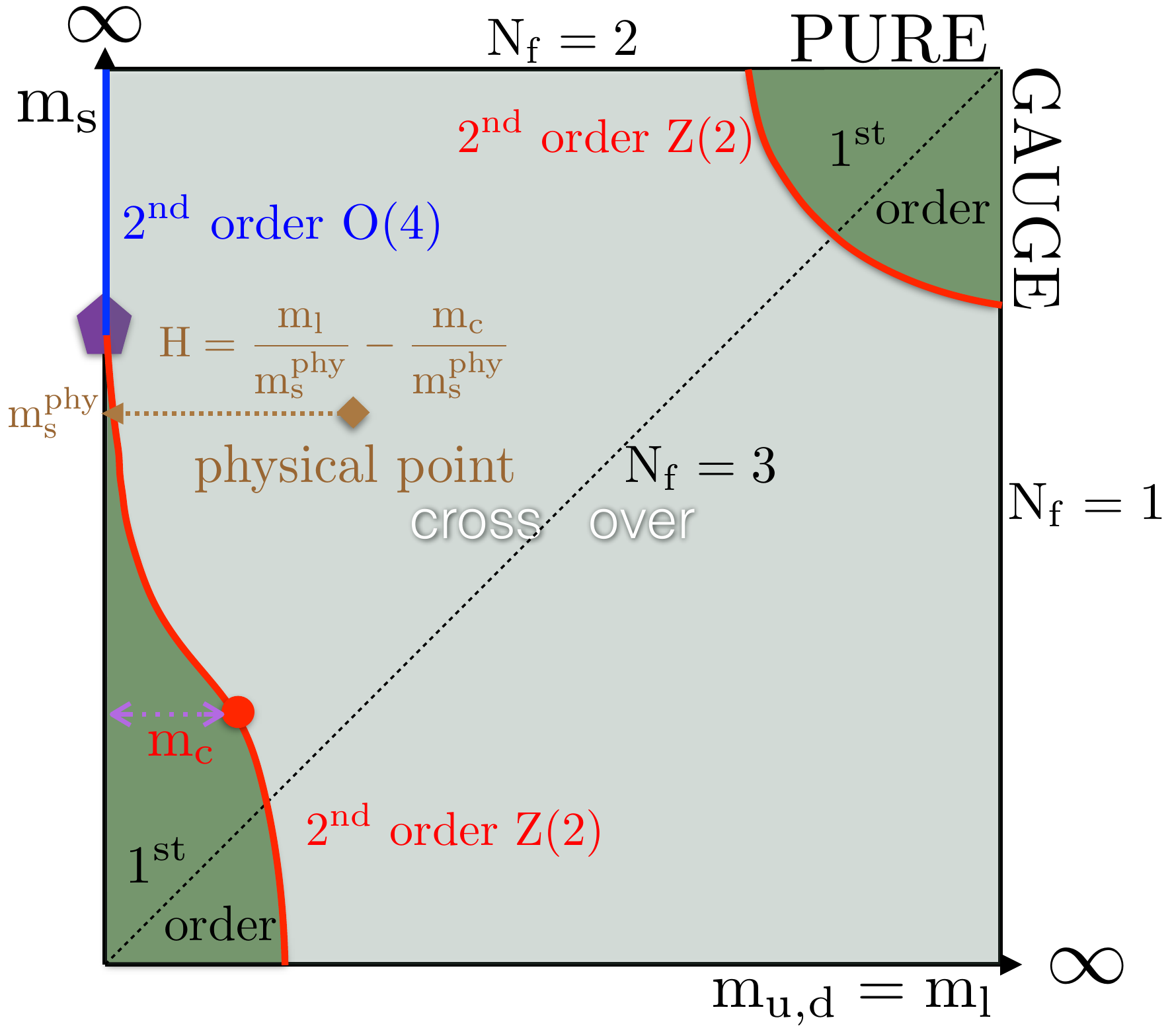}~
	\end{center}
	\caption{Schematic QCD phase structure with different values of quark masses ($m_{u,d},m_{s}$) at zero baryon number density for $m_{s}^{\mathrm{tri}}<m_{s}^{\mathrm{phy}}$ (left) and $m_{s}^{\mathrm{tri}}>m_{s}^{\mathrm{phy}}$ (right).}
	\label{Fig:latticedata}	
\end{figure}

One of the basic goals of lattice QCD calculations at non-zero temperature is to understand the QCD phase diagram \cite{Ding:2015ona}.
At zero baryon chemical potential, the QCD phase structure may depend on the number of light quark flavors \cite{PhysRevD.29.338} which is summarized in the Columbia plot in two scenarios as shown in Fig.~\ref{Fig:latticedata}. 
It is concluded that the physical point ($m_{u,d}^{\mathrm{phy}},m_{s}^{\mathrm{phy}}$) is located in the crossover region~\cite{PhysRevD.85.054503,Aoki:2006aa, Bazavov:2018mes}.
The first order phase transition regions and the crossover region are separated by second order phase transition lines which belong to the $Z(2)$ universality class. 
In the chiral limit of $N_{f}=2$ theory, if $U_{A}(1)$ symmetry remains broken at the chiral transition temperature, the chiral phase transition is a second order phase transition belonging to an $O(4)$ universality class~\cite{PhysRevD.29.338}.
Thus the chiral first order region in the left bottom corner of Columbia plot, the second order $O(4)$ line for $N_{f}=2$ case and the second order $Z(2)$ line are supposed to meet at a tri-critical point $m_{s}^{\mathrm{tri}}$. 
The location of the tri-critical point is still an open question. It is possible that the tri-critical point shifts to infinite strange quark mass~\cite{PhysRevD.93.114507}.
\vspace{-0.2cm}
The nature of the chiral phase transition at zero baryon chemical potential is also relevant for our understanding of the QCD phase diagram at non-zero chemical potential. 
If $m_{s}^{\mathrm{tri}}<m_{s}^{\mathrm{phy}}$, it is expected that in the chiral limit there will be a second order phase transition which belongs to the $O(4)$ universality class as seen from the Fig.~\ref{Fig:latticedata} (left). In this case, in the chiral limit, there might exist a tri-critical point as if QCD system becomes a first order phase transition in large baryon chemical potential.
If $m_{s}^{\mathrm{tri}}>m_{s}^{\mathrm{phy}}$, towards the chiral limit the system passes through the $Z(2)$ critical line to a first order phase transition region as shown in Fig.~\ref{Fig:latticedata} (right). In this case the chiral phase transition may be first order for all values of the chemical potential or, there may exist a critical point such that the transition becomes a crossover transition at large baryon chemical potential.

In this proceedings, we focus on the determination of the chiral phase transition temperature $T_{c}^{0}$ in the chiral limit and continuum limit, and we will also discuss the nature of the chiral phase transition. Previous studies have been reported in Ref.~\cite{Ding:2015pmg,Li:2017aki,Ding:2018auz}.


\section{Observables and definitions}
The universal behavior of the order parameter $M$ and its susceptibility $\chi_{M}$ can be described by the so-called Magnetic Equation of State (MEOS)~\cite{PhysRevD.80.094505} as follows
\begin{small}
	\begin{equation}
	M(t,h)=h^{1/\delta}f_{G}(z)\quad \mathrm{and}\quad
	\chi_M(t,h) = \frac{\partial M}{\partial H} =  h_{0}^{-1}h^{1/\delta-1}f_{\chi}(z).
	\label{eq.M}
	\end{equation}
\end{small}
Here $z=th^{-1/\beta\delta}$ is a scaling variable, $t=\frac{1}{t_{0}}\frac{T-T_{c}^{0}}{T_{c}^{0}}$ is the reduced temperature and $h=H/h_{0}=\frac{m_{l}}{m_{s}}/h_{0}$ is the symmetry breaking field. $\beta$, $\delta$ are universal critical exponents which are uinque for a given universality class as shown in Table~\ref{Univer}.

	\begin{table}[h]
		\begin{center}\small
			\begin{tabular}{|c|c|c|c|c|c|c|}
				\hline
				Model &$\beta$ & $\delta$ & $z_{p}$ & $z_{60}$& $f_{G}(z_{p})$ & $f_{\chi}(z_{p})$\\
				\hline
				$Z(2)$ & 0.3258&4.805 & 2.00(5)& 0.10(1) & 0.548(10)&0.3629(1)\\
				$O(2)$ & 0.349 &4.780 & 1.58(4) & -0.005(9) & 0.550(10)&0.3489(1) \\					
				$O(4)$ & 0.380 &4.824& 1.37(3) & -0.013(7) & 0.532(10)&0.3430(1)\\				
				\hline
			\end{tabular}
		\end{center}
		 \vspace{-0.4cm}
		\caption{Universal critical exponents $\beta$, $\delta$ for $Z(2)$, $O(2)$ and $O(4)$ 3-d universality classes. Also given is the peak location of $f_{\chi}$, i.e. $z_{p}$, and the location $z_{60}$ where the height of the $f_{\chi}$ is 60\% of its peak height and the values of $f_{G}(z_{p})$ and $f_{\chi}(z_{p})$. }
		\label{Univer}                   
	\end{table}
 \vspace{-0.3cm}
 
Three non-universal parameters $h_{0}$, $t_{0}$, $T_{c}^{0}$ are unique for a particular system, e.g. $T_{c}^{0}$ is the critical temperature of chiral phase transition in the light quark chiral limit.
For scaling variable $z_{X}$, it is related to a temperature $T_{X}$ as follows

\begin{small}
	\begin{equation}
T_{X}(H)=T_{c}^{0}+z_{X}T_{c}^{0}H^{1/\beta\delta}/z_{0},\quad z_{0}=h_{0}^{1/\beta\delta}/t_{0}.
	\label{eq.M1}
	\end{equation}
\end{small}
At the peak location of $f_{\chi}$, i.e. $z_{X}=z_{p}$, we have the relationship between pseudo-critical temperature $T_{\rm{pc}}$ and the critical temperature $T_{c}^{0}$, e.g. $T_{\rm{pc}}=T_{c}^{0} + z_{p}T_{c}^{0}H^{1/\beta\delta}/z_{0}$. Here we analyze two other estimators for the chiral phase transition temperature, defined by two specific values of the scaling variable $z$, i.e. $z_{60}$ and $z_{\delta}$. The former is defined by $f_{\chi}(z_{60})=0.6f_{\chi}(z_{p})$ with $z_{60}<z_{p}$ and the corresponding $T_{60}$ is defined as $\chi_{M}(T_{60})=0.6\chi_{M}(T_{\rm{pc}})$ with $T_{60}< T_{\rm{pc}}$.

\begin{figure}[ht]
	\begin{center}
		\includegraphics[width=0.41\linewidth]{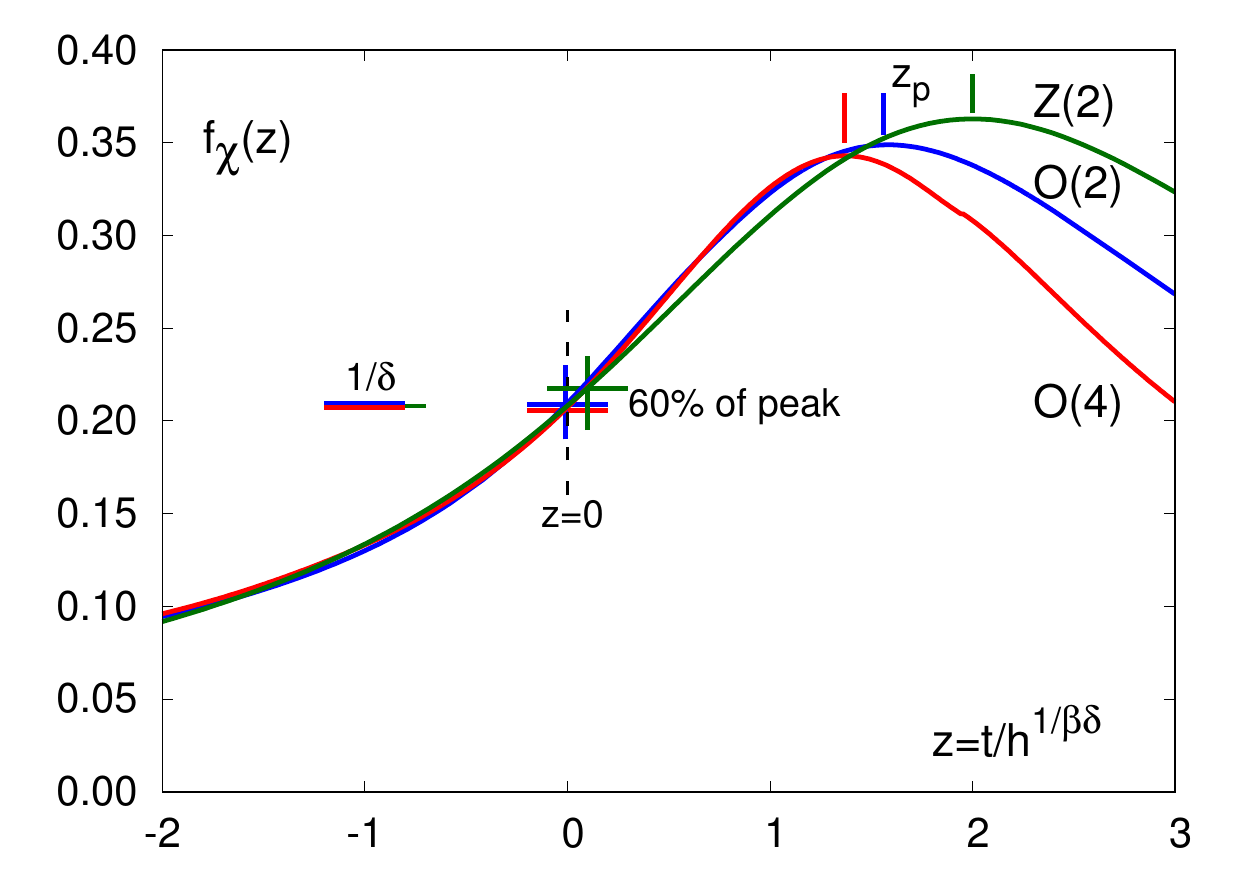}
		\includegraphics[width=0.41\linewidth]{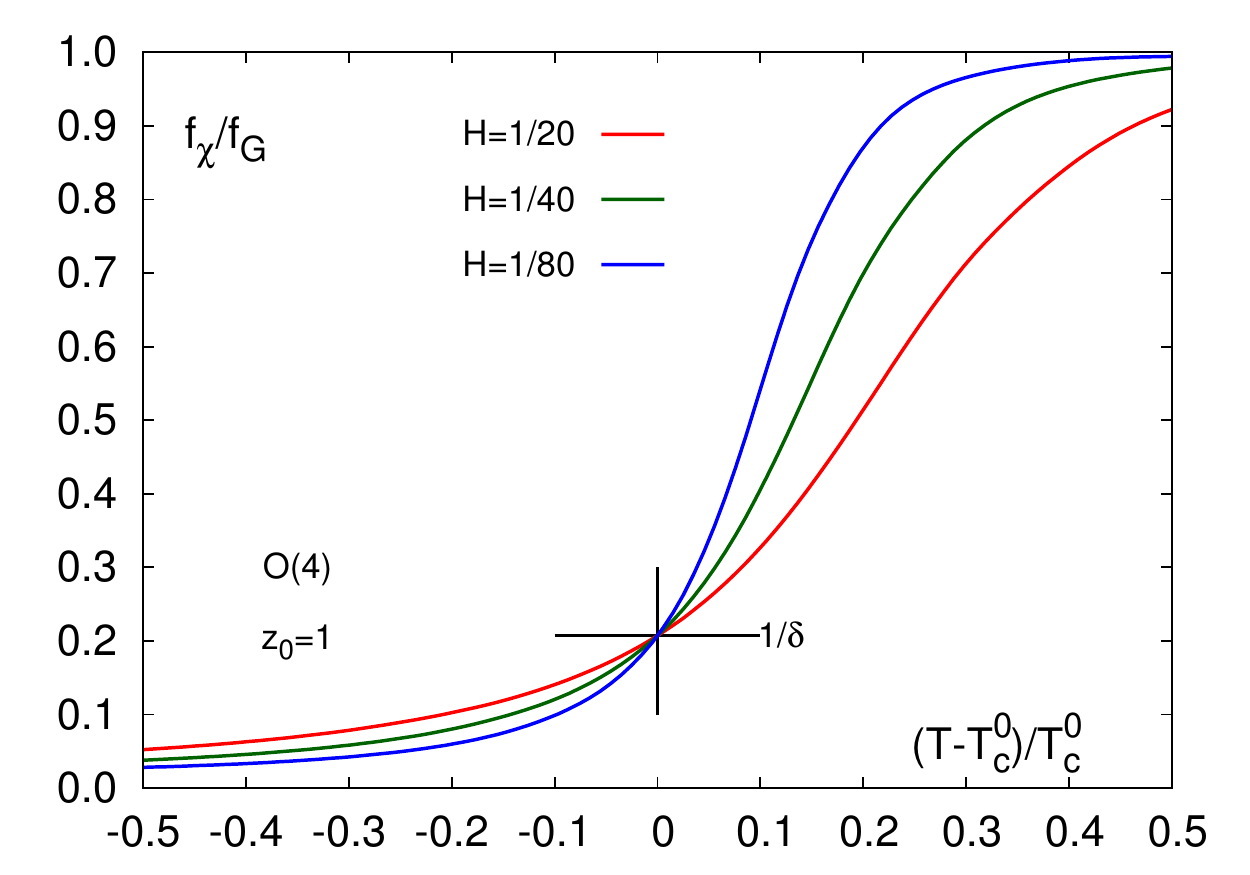}		
	\end{center}
			 \vspace{-0.4cm}
	\caption{
		Left: Scaling function $f_{\chi}$ for $O(4)$, $O(2)$ and $Z(2)$ universality classes, $z_{p}$ is the peak position of $f_{\chi}$. $z_{60}\approx 0$ for these universality classes. Right: Ratio of scaling functions using $O(4)$ exponents for three different values of $H$~\cite{Ding:2018auz}. 
	}
	\label{Fig:data_sets1}
\end{figure}

 Since $z_{60}$ is very close to zero the $H$-dependent term In Eq.~\ref{eq.M1} is suppressed by at least by an order of magnitude compared to $z_{p}$. This is shown in the left panel of Fig.~\ref{Fig:data_sets1} and Table~\ref{Univer}, for relevant universality classes. We thus can estimate $T_{c}^{0}$ by investigating the values of $T_{60}$ as follows
\begin{small}
	\begin{equation}
	T_{60}(H)=T_{c}^{0} + z_{60}T_{c}^{0}H^{1/\beta\delta}/z_{0},
	\label{eq.M2}
	\end{equation}
\end{small}

In the right panel of Fig.\ref{Fig:data_sets1}, we plot $f_{\chi}/f_{G}$ vs. $(T-T_{c}^{0})/T_{c}^{0}$ for $O(4)$ universality class where we have set $z_{0}$=1 for simplicity. The different curves, corresponding to different $H$, meet at a unique crossing point $(0, 1/\delta)$~\cite{PhysRevD.50.6954}. This thus drives us to estimate $T_{c}^{0}$ by looking at $H\chi_{M}/M$,
\begin{small}
	\begin{equation}
 \frac{H\chi_{M}(T_{\delta}, V, H)}{M(T_{\delta}, V, H)}=\frac{1}{\delta}\Rightarrow T_{c}^{0}=\lim_{H\to 0}\lim_{V\to \infty} T_{\delta}(V,H).
	\label{eq.M3}
	\end{equation}
\end{small}
As shown in Eq.~\ref{eq.M3}, $T_{c}^{0}$ can be estimated by looking at $T_{\delta}(V,H)$ in the infinite volume limit and chiral limit.

By looking at following equation we will be able to investigate the nature of chiral phase transition
\begin{small}
	\begin{equation}
 \frac{M}{\chi_{M}}=(H-H_{c})\frac{f_{G}(z)}{f_{\chi}(z)}.
	\label{eq.M4}
	\end{equation}
\end{small}

 In the case of $m_{s}^{\rm{tri}} < m_{s}^{\rm{phy} } $, $H_{c}$ is zero and the chiral phase transition is a second order phase transition in the chiral limit, the universality class is expected to be $O(4)$. In the case of $m_{s}^{\rm{phy} } < m_{s}^{\rm{tri}}$, $H_{c}$ is nonzero and the chiral phase transition is a first order phase transition in the chiral limit, where the corresponding universality class of the second order phase transition occuring at some $H_{c}>0$ is $Z(2)$. 
 Since $f_G(z)/f_{\chi}(z)$ at $z\simeq 0$ and $z_{p}$ is a number fixed by universality class, one can study the order of the chiral phase transition through the relation between $M/\chi_{M}$ and $H$ at $T_{60}$ and $T_{\rm{pc}}$.
\section{Lattice setup}
In our simulations of (2 + 1)-flavor QCD we have used Highly Improved Staggered Quarks and tree-level improved gauge action (HISQ/tree).
The strange quark mass is chosen to its physical quark mass value $m_{s}^{\mathrm{phy}}$, and the light quark masses values are varied from $m_{s}^{\mathrm{phy}}/160$ to $m_{s}^{\mathrm{phy}}/20$ which correspond to 55 $\rm{MeV}$ $\leq$ $m_{\pi} \leq $ 160 $\rm{MeV}$. To perform the continuum limit, the temporal extent $N_{\tau}$ is taken to be 6, 8 and 12 and the spatial volumes used are in the range 4 $\leq$ $N_{\sigma}/N_{\tau}$ $ \leq$ 8.

As shown in the Table~\ref{T6_8_12}, for each data set we have performed at least 10000 time units (TUs) at each temperature, where gauge configurations are separated by every 5 TUs.
We used 50 random noise vectors on each gauge field configuration and constructed unbiased estimators for the various traces to compute the chiral condensate and its susceptibility.
\begin{table}[htbp]
	\scriptsize
	\begin{center}
		\begin{tabular}{|cc|c|}
			\hline
			$N_{\sigma}^{3}\times N_{\tau}$& $ \frac{m_{l}}{m_{s}^{\rm phy}}$    & average \# of TU \\ \hline
			$24^{3}\times6$   & 1/20  &  23000\\  
			$24^{3}\times6$   & 1/27  &  13800\\  
			$32^{3}\times6$   & 1/40  &  20000\\  
			$40^{3}\times6$   & 1/60  &  15000\\  
			$24^{3}\times6$   & 1/80  &  40000\\  
			$32^{3}\times6$   & 1/80  &  26000\\  
			$48^{3}\times6$   & 1/80  &  10000\\  
			\hline
		\end{tabular}
		\begin{tabular}{|cc|c|}
			\hline
			$N_{\sigma}^{3}\times N_{\tau}$& $ \frac{m_{l}}{m_{s}^{\rm phy}}$     & average \# of TU \\ \hline
			$24^{3}\times8$   & 1/40  &  100000\\  
			$32^{3}\times8$   & 1/40  &  32000\\  
			$40^{3}\times8$   & 1/40  &  14000\\  
			$32^{3}\times8$   & 1/80  &  80000\\  
			$40^{3}\times8$   & 1/80  &  35000\\  
			$56^{3}\times8$   & 1/80  &  20000\\  
			$56^{3}\times8$   & 1/160 &  14000\\  	
			\hline
		\end{tabular}
		\begin{tabular}{|cc|c|}
			\hline
			$N_{\sigma}^{3}\times N_{\tau}$& $ \frac{m_{l}}{m_{s}^{\rm phy}}$     & average \# of TU \\ \hline
			$42^{3}\times12$   & 1/40  &  50000\\  
			$60^{3}\times12$   & 1/40  &  32000\\  
			$48^{3}\times12$   & 1/80  &  37000\\  
			$60^{3}\times12$   & 1/80  &  18000\\  
			$72^{3}\times12$   & 1/80  &  15000\\  
			\hline
		\end{tabular}
		
	\end{center}
	\vspace{-0.4cm}
	\caption{Current statistics for $N_{\tau}$ = 6, 8 and 12 lattices.}	
	\label{T6_8_12}
\end{table}%

\section{Results}
We study the subtracted chiral order parameter $M$ and its susceptibility $\chi_{M}$. To avoid the distorsion of the temperature dependence at low temperatures of the chiral order parameter we use the following
definitions
\begin{small}
	\begin{equation}
	M={m_{s}}\left(\left\langle\bar{\psi}\psi\right\rangle_{l} -\frac{2m_{l}}{m_{s}}\left\langle\bar{\psi}\psi\right\rangle_{s}\right)/f_{K}^{4},\quad f_{K}=(156.1/\sqrt{2})\ \rm{MeV},
	\label{eq.sub_pbp}
	\end{equation}
\end{small}
\begin{small}
	\begin{equation}
	\chi_{M}\equiv\frac{\partial M}{\partial H}  \equiv m_{s}^2\chi_{l,\mathrm{subtot}}/f_{K}^{4}, \quad  \chi_{l,\mathrm{subtot}}=\frac{\partial }{\partial m_{l}}\left(\left\langle\bar{\psi}\psi\right\rangle_{l} -\frac{2m_{l}}{m_{s}}\left\langle\bar{\psi}\psi\right\rangle_{s}\right).
	\label{eq.sub_pbp2}
	\end{equation}
\end{small}
We replace the factors of $T^4$ by the appropriate factors of the kaon decay constant, $f_{K}^4$.
\begin{figure}[ht]
	\begin{center}
		\includegraphics[width=0.4\linewidth]{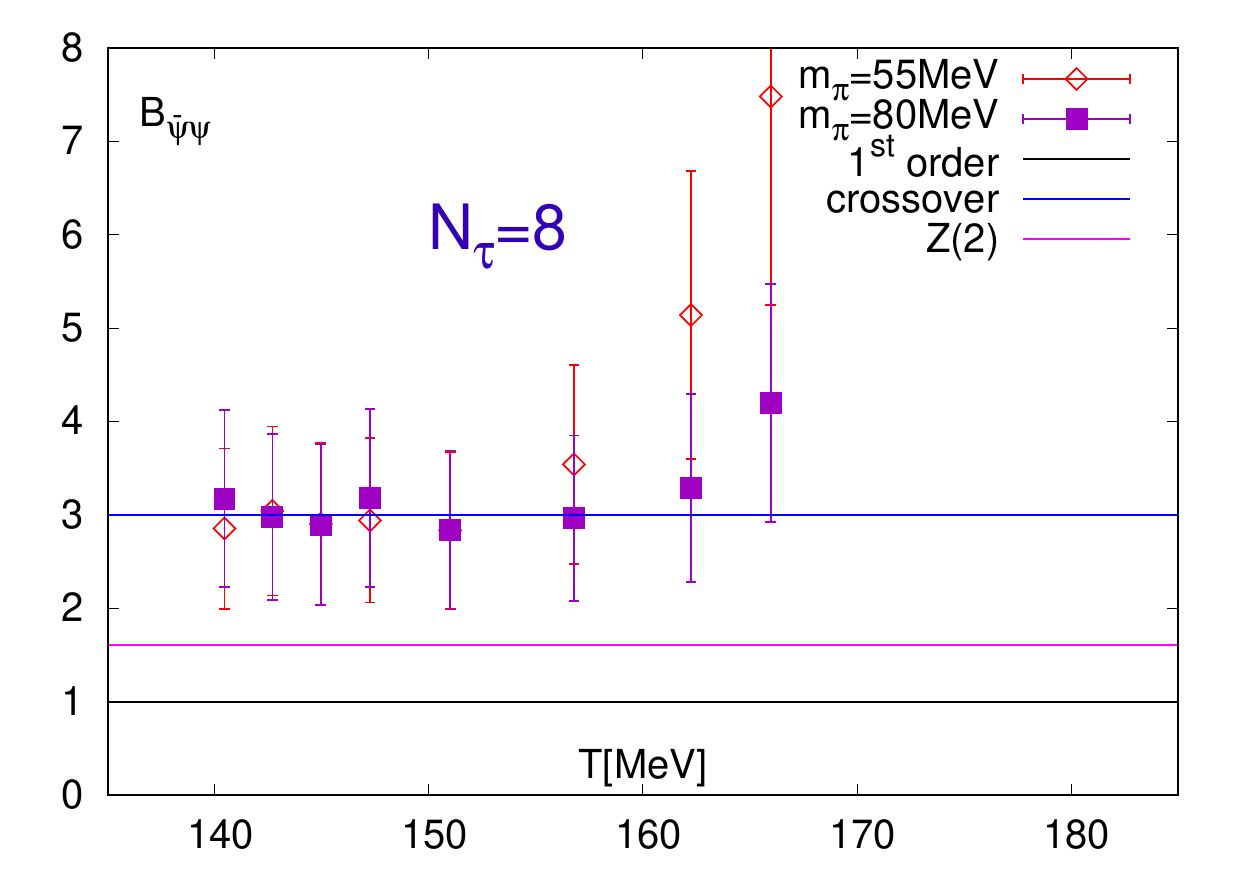}
		\includegraphics[width=0.4\linewidth]{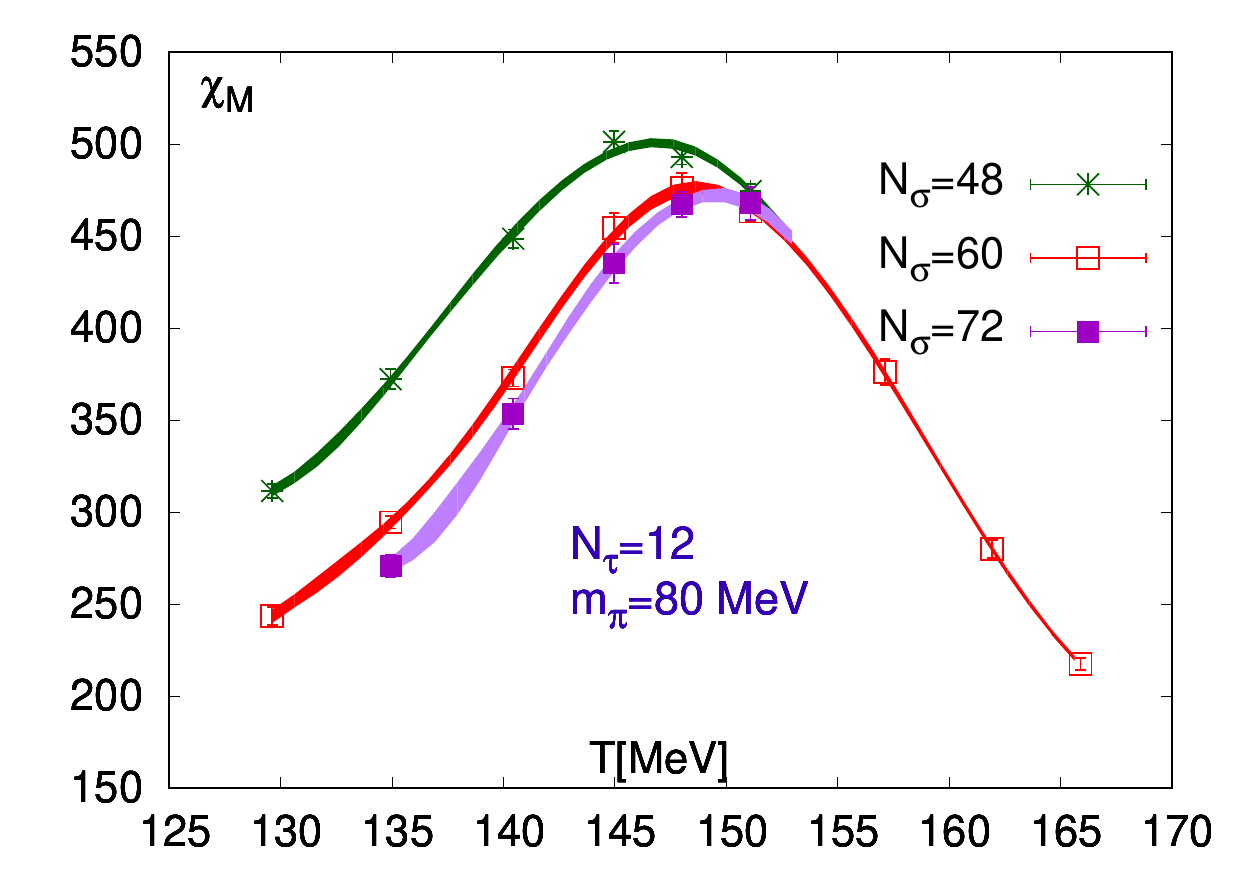}
	\end{center}
	\vspace{-0.4cm}
	\caption{
		Left: Binder cumulant of chiral condensate on $N_{\tau}=8$ lattices for $m_{s}/m_{l}=80$ and 160. Right: Volume dependence of chiral susceptibilities on $N_{\tau}=12$ lattices for $m_{s}/m_{l}=80$. 
	}
	\label{Fig:data_sets}
\end{figure}

In the left panel of Fig.~\ref{Fig:data_sets} we show the Binder cumulant of chiral condensate on $N_{\tau}=8$ lattices for $m_{\pi}=$ 80 and 55 MeV. Here the Binder cumulant is defined as $B_{X}=\left\langle (X -\left\langle X \right\rangle )^4 \right\rangle /\left\langle (X -\left\langle X \right\rangle )^2 \right\rangle^{2}$. The plot shows that there is no evidence of first order phase transition in our current pion mass window 55 $\rm{MeV}$ $\leq$ $m_{\pi} \leq $ 160 $\rm{MeV}$. Also as seen from the right panel of Fig.~\ref{Fig:data_sets}, chiral susceptibilities obtained on $N_{\tau}=12$ lattices do not grow linearly with the volume which implies that there is no first order phase transition in our current pion mass window. We also observe that $T_{\rm{pc}}$ is larger for larger volume.
\begin{figure}[htbp]			
	\begin{center}
\includegraphics[width=0.320\textwidth]{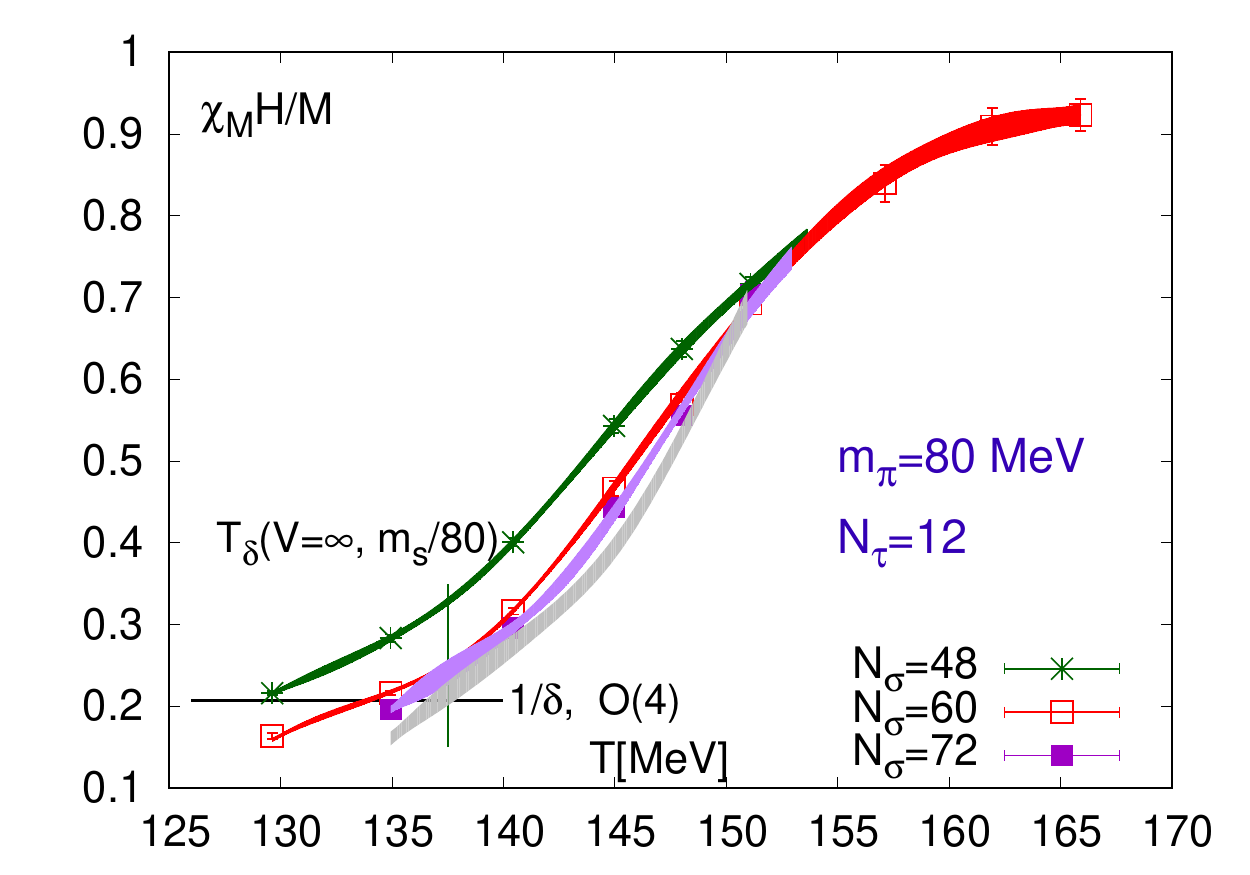}	
\includegraphics[width=0.320\textwidth]{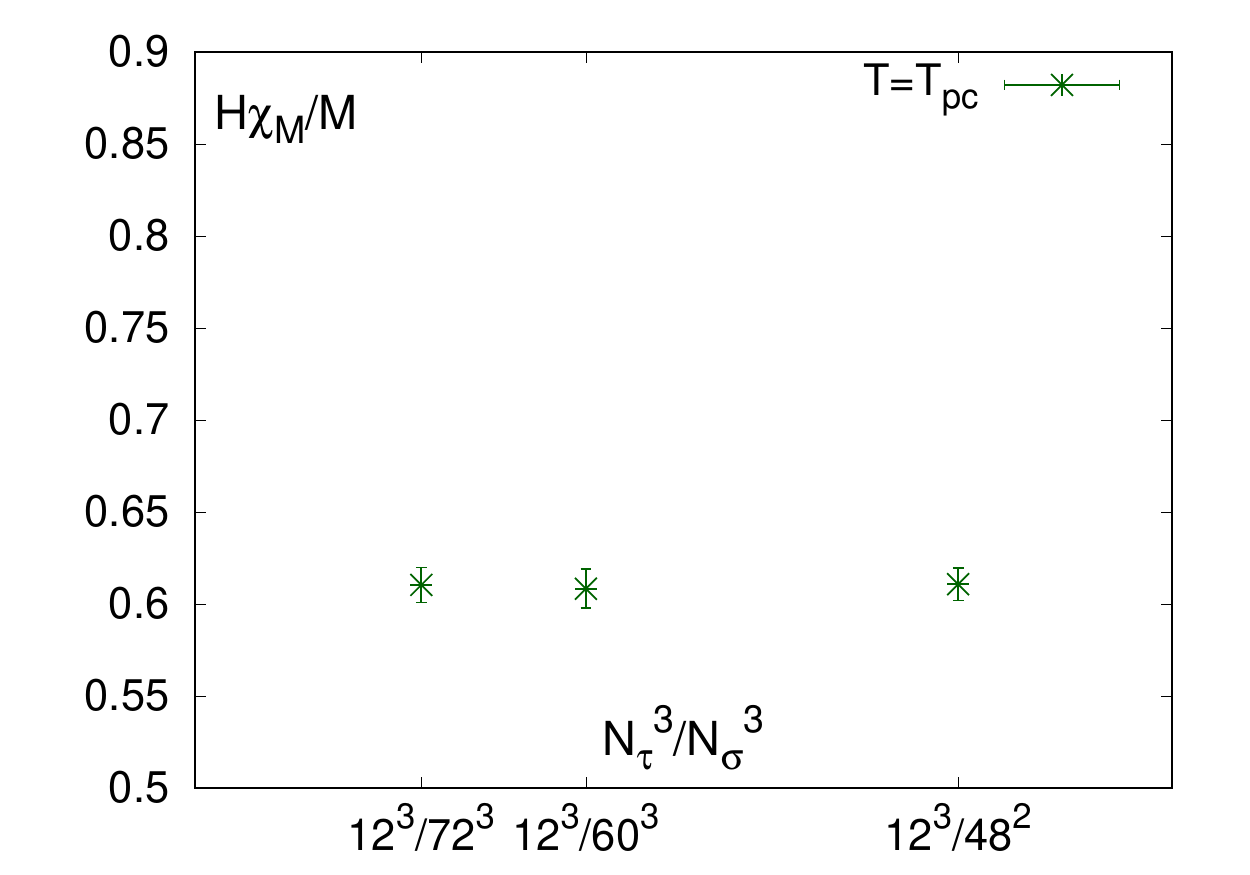}~	
		\includegraphics[width=0.320\textwidth]{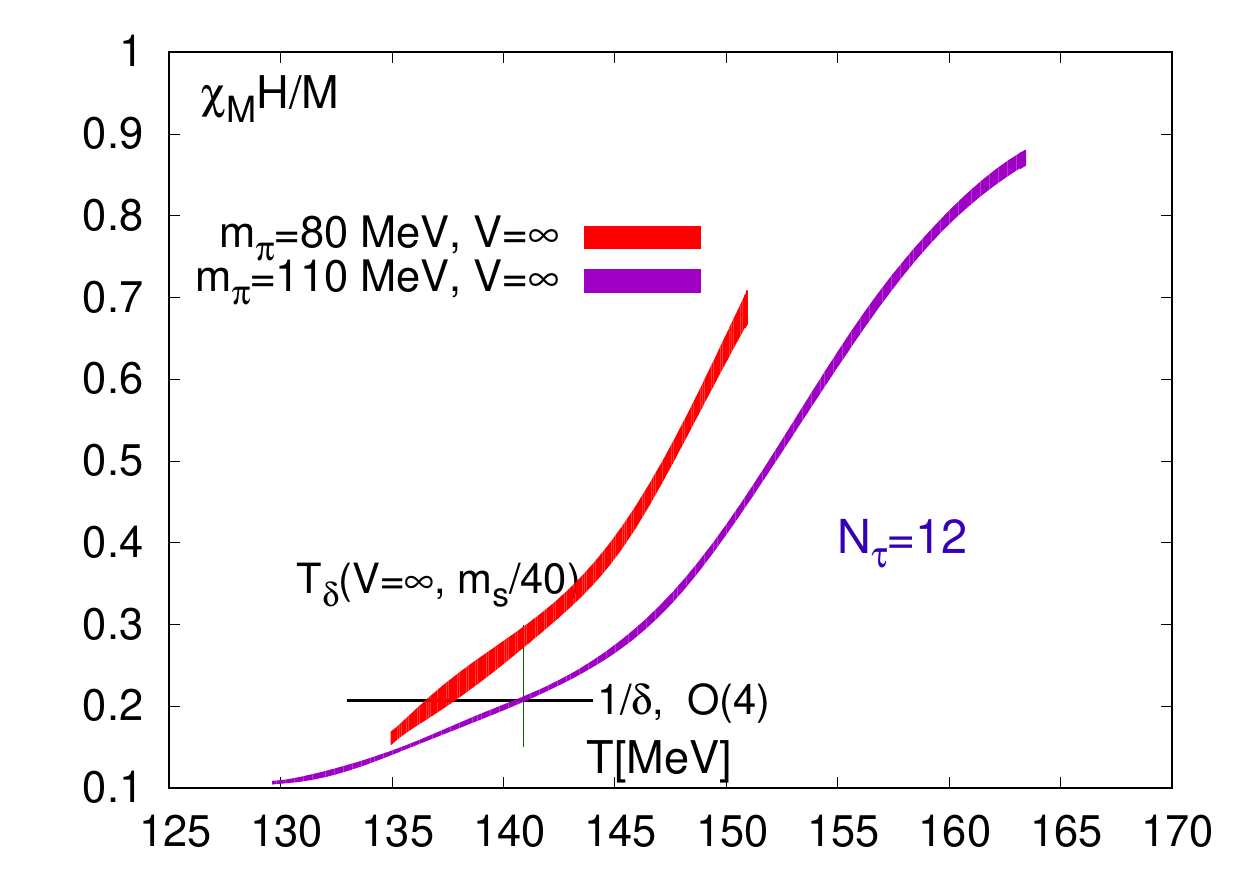}~
	\end{center}	
	\vspace{-0.7cm}
	\caption{Volume dependence of $\chi_{M}H/M$ on $N_{\tau}=$12 lattices.		}
	\label{fg.V}
\end{figure}

We then show the volume dependence of $\chi_{M}H/M$ in Fig.~\ref{fg.V}. At a fixed temperature $H\chi_{M}/M$ becomes smaller in larger volume, and for a fixed volume it increases with increasing temperature. $H\chi_{M}/M$ at $T_{\rm{pc}}$ is almost volume independent as shown in the middle plot of Fig.~\ref{fg.V}. Similar results are also obtained from $N_{\tau}=12$ lattices with $m_{l}=m_{s}/40$ and $N_{\tau}=8$ lattices with $m_{l}=m_{s}/80$. As shown in the left plot of Fig.~\ref{fg.V}, $T_{\delta}(V,H)$ increases with the increasing volume. Thus, we performed $1/V$ extrapolation as represented by the grey band. This gives $T_{\delta}(V\to\infty, m_{l}=m_{s}/80)\approx$ 138 MeV. Similar analyses are done for $m_{s}/40$ which gives $T_{\delta}(V\to\infty, m_{l}=m_{s}/40)\approx$ 141 MeV as shown as the vertical line in the right plot of Fig.~\ref{fg.V}. This figure also shows that results for different quark masses do not cross at a unique point $(T_{c},1/\delta)$ as one would expect in the scaling regime (see Fig.~\ref{Fig:data_sets1}(right)). This reflects the importance of regular contributions. We then take the chiral limit by
performing linear extrapolation in $H$, which gives an estimate $T_{c}^{0}(N_{\tau}=12)\simeq 134(2) $ MeV. 

\begin{figure}[h]
	\begin{center}
\includegraphics[scale=0.41]{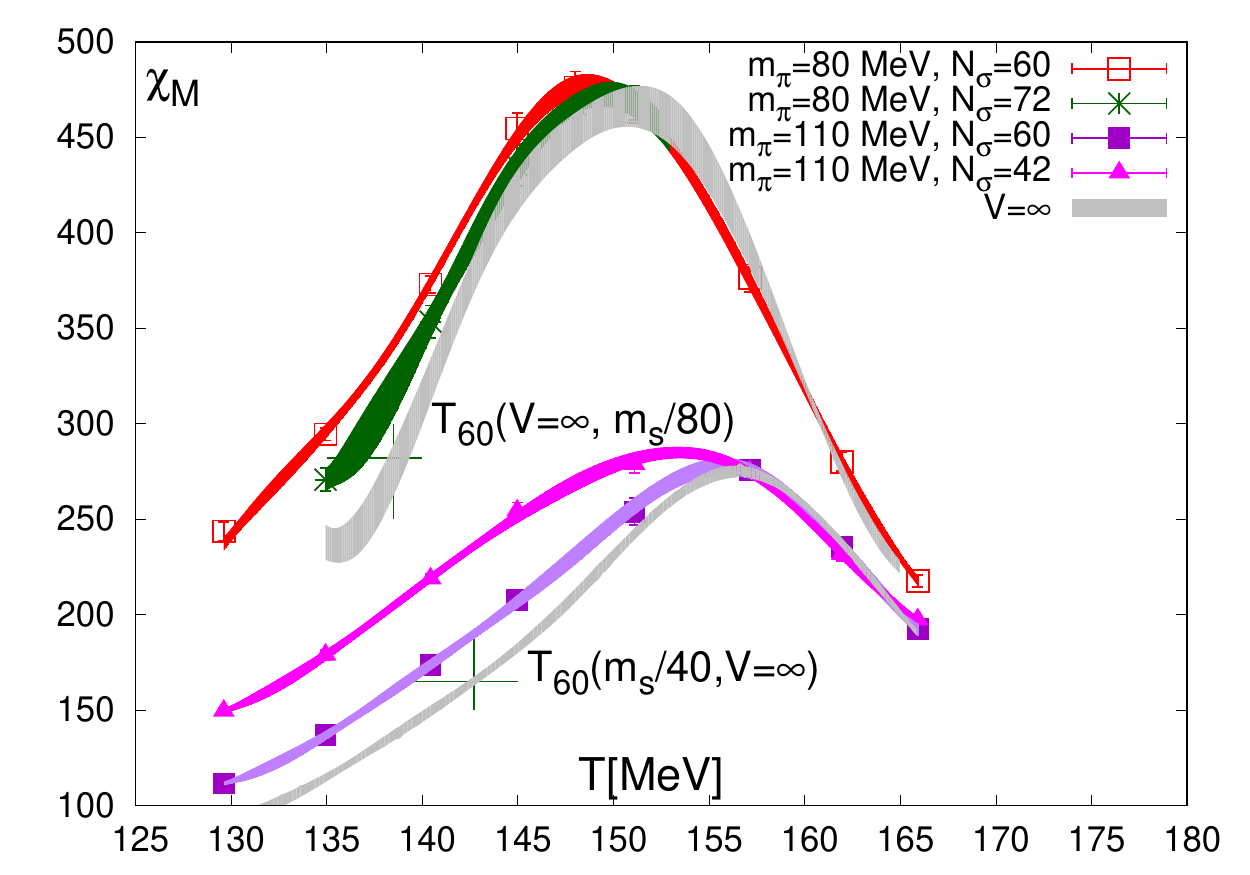}~		
\includegraphics[scale=0.41]{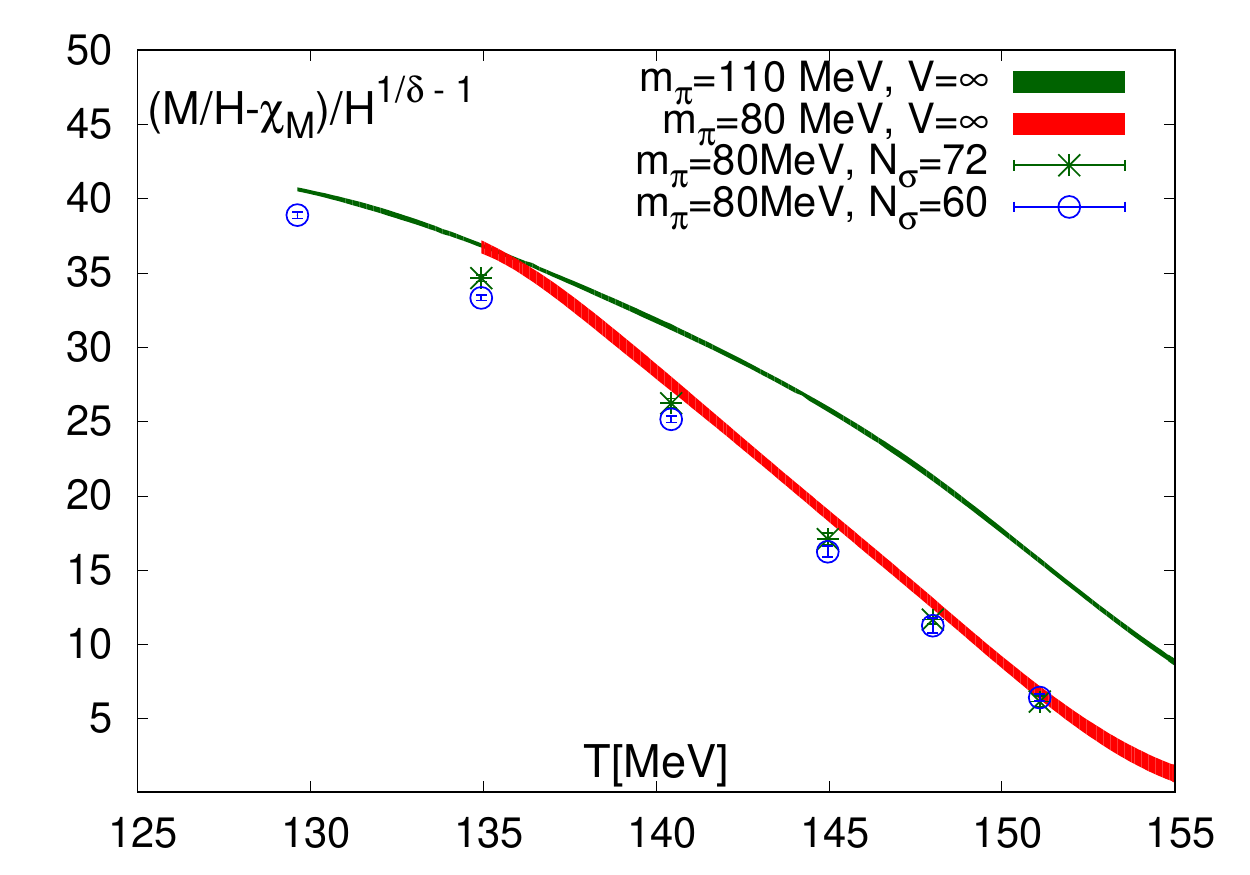}

	\end{center}
\vskip -0.3 in
\caption{Left: $T_{60}$ for $N_{\tau}=12$ lattices.
	Right: $T_{c}^{0}$ estimated by the crossing point of $M/H-\chi_{M}$ rescaled by $H^{1/\delta-1}$}
\label{fg.T60Tdelta}
\end{figure}

\vspace{-0.2cm}
In the left plot of Fig.~\ref{fg.T60Tdelta} we show the extraction of $T_{60}$ on $N_{\tau}=12$ lattices. We performed the $1/V$ extrapolation as shown as the grey bands, and then performed linear chiral extrapolation which gives a consistent result $T_{c}^{0}\approx$ 134(2) MeV.
A sanity check for the $T_{c}^{0}$ is to look at the crossing point of $M/H-\chi_{M}$ rescaled by $H^{1/\delta-1}$. The advantage of $M/H-\chi_{M}$ is that it removed the regular contributions linear in $H$. As shown in the right plot of Fig.~\ref{fg.T60Tdelta}, $(M/H-\chi_{M})/H^{1/\delta-1}$ in the infinite volume limit with $H=1/40$ (the green band) and $H=1/80$ (the red band) meet at a "crossing point" which is roughly around 135 MeV. Similar procedures e.g. ($T_{\delta},\ T_{60},\ M/H-\chi_{M}$) are done for $N_{\tau}=8$ and 6 which gives $T_{c}^{0}=$ 142(2) and 147(2) MeV, respectively.
Rather than doing $1/V$ and linear in $H$ extrapolations we also analyze the finite size scaling using $O(4)$ scaling functions, 
and the continuum extrapolated $T_{c}$ values are systematically lower by 2-3 MeV which is one source of our systematic uncertainty.
The continuum extrapolations discarding results obtained on $N_{\tau}=6$ lattices gives about 3 MeV lower $T_{c}^{0}$, this is another source of our systematic uncertainty.

\begin{figure}[h]
	\begin{center}
		\includegraphics[scale=0.43]{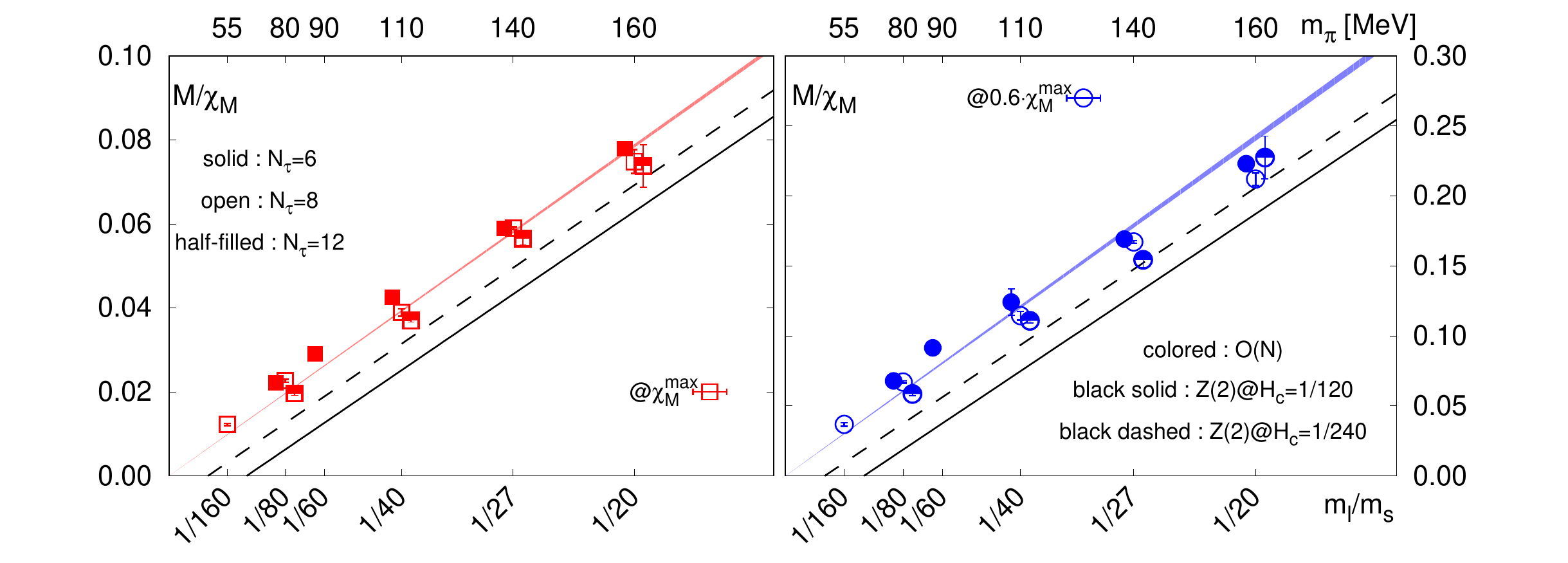}
	\end{center}
	\vskip -0.3 in
	\caption{$M/\chi_{M}$ is plotted for different $N_{\tau}$ along with the scaling
		expectations from different universality classes.}
	\label{fg.MtochiMratio}
\end{figure}

 To study the nature the chiral phase transition, we look at the ratio $M/\chi_{M}$ (c.f. Eq.~\ref{eq.M4}). We show the quark mass dependence of $M/\chi_{M}$ at $T_{\rm{pc}}$ and $T_{60}$ in the left plot and right plot of Fig.~\ref{fg.MtochiMratio}, respectively. As discussed before, $H\chi_{M}/M$ at $T_{\rm{pc}}$ is almost volume independent, and this indicates that all the data points shown in the Fig.~\ref{fg.MtochiMratio} can be regarded as being in the infinite volume limit. The colored band in the plots represents the difference between $O(2)$ and $O(4)$ universality classes. 
In Fig.~\ref{fg.MtochiMratio} we also compare our lattice results on the quark mass dependence of $M/\chi_M$ with the scenarios of Z(2) phase transition with non-zero $H_c$ corresponding to $m_l/m_s=1/120$ and $m_l/m_s=1/240$. As one can see from the figure our lattice results are way above these expectations. Thus, if there is a first order phase transition in the chiral limit of (2 + 1)-flavor QCD, it should happen for quark masses smaller than $m_s/160$.
\vspace{-0.2cm}
\section{Summary}
We have performed lattice simulations of (2 + 1)-flavor QCD using HISQ/tree action. To study the chiral phase transition temperature $T_{c}^{0}$ in the chiral \& continuum limit, the light quark mass window was chosen to be $m_{s}^{\rm phy}/160 \leq m_{l} \leq m_{s}^{\rm phy}/20 $, which correspond to the pion mass window 55 MeV$\leq m_{\pi} \leq$160 MeV, $N_{\tau}$ was set to 6, 8, and 12, and the corresponding $N_{\sigma}$ ranging from 4$N_{\tau}$ to 8$N_{\tau}$. The current estimates of $T_{c}^{0}$ on $N_{\tau}=$ 6, 8 and 12 lattices are 147(2) MeV, 142(2) MeV and 135(3) MeV, respectively. Including all systematic uncertainties, our current estimate of $T_{c}^{0}$ in the continuum limit is $T_{c}^{0}=132_{-6}^{+3}$ MeV~\cite{Ding:2019prx}. By looking at the ratio $M/\chi_{M}$ as a function of light quark mass, the chiral phase transition is more like a second order phase transition instead of a first order phase transition.

\label{sc.acknowledgments}
\vspace*{-0.3 cm}

\bibliographystyle{JHEP}
\bibliography{lattice2018.bib}

\providecommand{\href}[2]{#2}\begingroup\raggedright\begin{thebibliography}{10}

\bibitem{Ding:2015ona}
H.-T. Ding, F.~Karsch and S.~Mukherjee, \emph{{Thermodynamics of
  strong-interaction matter from Lattice QCD}},
  \href{https://doi.org/10.1142/S0218301315300076}{\emph{Int. J. Mod. Phys.}
  {\bfseries E24} (2015) 1530007}
  [\href{https://arxiv.org/abs/1504.05274}{{\ttfamily 1504.05274}}].

\bibitem{PhysRevD.29.338}
R.~D. Pisarski and F.~Wilczek, \emph{Remarks on the chiral phase transition in
  chromodynamics}, \href{https://doi.org/10.1103/PhysRevD.29.338}{\emph{Phys.
  Rev. D} {\bfseries 29} (1984) 338}.

\bibitem{PhysRevD.85.054503}
{\scshape HotQCD Collaboration} collaboration, A.~Bazavov, T.~Bhattacharya,
  M.~Cheng, C.~DeTar, H.-T. Ding, S.~Gottlieb et~al., \emph{Chiral and
  deconfinement aspects of the qcd transition},
  \href{https://doi.org/10.1103/PhysRevD.85.054503}{\emph{Phys. Rev. D}
  {\bfseries 85} (2012) 054503}.

\bibitem{Aoki:2006aa}
Y.~Aoki, G.~Endr{\H o}di, Z.~Fodor, S.~D. Katz and K.~K. Szab{\'o}, \emph{The
  order of the quantum chromodynamics transition predicted by the standard
  model of particle physics}, {\emph{Nature} {\bfseries 443} (2006) 675 EP }.

\bibitem{Bazavov:2018mes}
A.~Bazavov et~al., \emph{{Chiral crossover in QCD at zero and non-zero chemical
  potentials}},  \href{https://arxiv.org/abs/1812.08235}{{\ttfamily
  1812.08235}}.

\bibitem{PhysRevD.93.114507}
O.~Philipsen and C.~Pinke, \emph{${N}_{f}=2$ qcd chiral phase transition with
  wilson fermions at zero and imaginary chemical potential},
  \href{https://doi.org/10.1103/PhysRevD.93.114507}{\emph{Phys. Rev. D}
  {\bfseries 93} (2016) 114507}.

\bibitem{Ding:2015pmg}
{\scshape Bielefeld-BNL-CCNU} collaboration, H.-T. Ding and P.~Hegde,
  \emph{{Chiral phase transition of $N_f$=2+1 and 3 QCD at vanishing baryon
  chemical potential}}, \href{https://doi.org/10.22323/1.251.0161}{\emph{PoS}
  {\bfseries LATTICE2015} (2016) 161}
  [\href{https://arxiv.org/abs/1511.00553}{{\ttfamily 1511.00553}}].

\bibitem{Li:2017aki}
S.-T. Li and H.-T. Ding, \emph{{Chiral phase transition of (2 + 1)-flavor QCD
  on $N_{\tau} = 6$ lattices}},
  \href{https://doi.org/10.22323/1.256.0372}{\emph{PoS} {\bfseries LATTICE2016}
  (2017) 372} [\href{https://arxiv.org/abs/1702.01294}{{\ttfamily
  1702.01294}}].

\bibitem{Ding:2018auz}
H.~T. Ding, P.~Hegde, F.~Karsch, A.~Lahiri, S.~T. Li, S.~Mukherjee et~al.,
  \emph{{Chiral phase transition of (2+1)-flavor QCD}},
  \href{https://doi.org/10.1016/j.nuclphysa.2018.10.032}{\emph{Nucl. Phys.}
  {\bfseries A982} (2019) 211}
  [\href{https://arxiv.org/abs/1807.05727}{{\ttfamily 1807.05727}}].

\bibitem{PhysRevD.80.094505}
S.~Ejiri, F.~Karsch, E.~Laermann, C.~Miao, S.~Mukherjee, P.~Petreczky et~al.,
  \emph{Magnetic equation of state in ($2+1$)-flavor qcd},
  \href{https://doi.org/10.1103/PhysRevD.80.094505}{\emph{Phys. Rev. D}
  {\bfseries 80} (2009) 094505}.

\bibitem{PhysRevD.50.6954}
F.~Karsch and E.~Laermann, \emph{Susceptibilities, the specific heat, and a
  cumulant in two-flavor qcd},
  \href{https://doi.org/10.1103/PhysRevD.50.6954}{\emph{Phys. Rev. D}
  {\bfseries 50} (1994) 6954}.

\bibitem{Ding:2019prx}
H.~T. Ding et~al., \emph{{The chiral phase transition temperature in
  (2+1)-flavor QCD}},  \href{https://arxiv.org/abs/1903.04801}{{\ttfamily
  1903.04801}}.

\end{thebibliography}\endgroup

\end{document}